\documentclass[fleqn,usenatbib]{mnras}

\usepackage{newtxtext,newtxmath}
\usepackage{wrapfig}
\usepackage{xcolor} 
\usepackage{multirow}

\usepackage[T1]{fontenc}

\DeclareRobustCommand{\VAN}[3]{#2}
\let\VANthebibliography\thebibliography
\def\thebibliography{\DeclareRobustCommand{\VAN}[3]{##3}\VANthebibliography}

\usepackage{graphicx}	
\usepackage{amsmath}	
\usepackage{deluxetable}
\usepackage{multirow}

\newcommand{\angstrom}{\mbox{\,\small\AA}}

\title[Environments of $z>5$ metals]{The environments and hosts of metal absorption at $z>5$}

\author[C. Doughty et al.]{
Caitlin C. Doughty,$^{1,2}$\thanks{E-mail: doughty@strw.leidenuniv.nl (CCD)}
Kristian M. Finlator,$^{2,3}$
\\
$^{1}$Leiden Observatory, Leiden University, P.O. Box 9513, 2300 RA Leiden, The Netherlands\\
$^{2}$Department of Astronomy, New Mexico State University, Las Cruces, NM 88003, USA \\
$^{3}$Cosmic Dawn Center (DAWN), Niels Bohr Institute, University of Copenhagen/DTU-Space, Technical University of Denmark
}

\date{Accepted XXX. Received YYY; in original form ZZZ}

\pubyear{2022}

\begin{document}
\label{firstpage}
\pagerange{\pageref{firstpage}--\pageref{lastpage}}
\maketitle

\begin{abstract}
A growing population of metal absorbers are observed at $z>5$, many showing strong evolution in incidence approaching the epoch of hydrogen reionization. Follow-up surveys examining fields around these metals have resulted in galaxy detections but the direct physical relationship between the detected galaxies and absorbers is unclear. Upcoming observations will illuminate this galaxy-absorber relationship, but the theoretical framework for interpreting these observations is lacking. To inform future $z>5$ studies, we define the expected relationship between metals and galaxies using the \emph{Technicolor Dawn} simulation to model metal absorption from $z=$ 5 \textendash 7, encompassing the end of reionization. We find that metal absorber types and strengths are slightly better associated with their environment than with the traits of their host galaxies, as absorption system strengths are more strongly correlated with the local galaxy overdensity than the stellar mass of their host galaxy. For redshifts prior to the end of the epoch of reionization, strong high ionization transitions like C IV are more spatially correlated with brighter galaxies on scales of a few hundred proper kpc than are low ionization systems, due to the former's preference for environments with higher UVB amplitudes and those ions' relative rarity at $z>6$. Post-reionization, the galaxy counts near these high-ionization ions are reduced, and increase surrounding certain low-ionization ions due to a combination of their relative abundances and preferred denser gas phase. We conclude that galaxy-absorber relationships are expected to evolve rapidly such that high-ionization absorbers are better tracers of galaxies pre-reionization while low-ionization absorbers are better post-reionization.
\end{abstract}

\begin{keywords}
galaxies: evolution -- galaxies: formation -- galaxies: high-redshift -- intergalactic medium -- quasars: absorption lines
\end{keywords}



\section{Introduction}
Prior to the formation of the first stars gas is pristine and devoid of heavier elements, but once star formation begins at $z\sim20$ heavier metals are formed via various nucleosynthetic pathways. Stellar feedback drives galactic winds which then distribute these metals within the virial radius and beyond, enriching the circumgalactic medium (CGM) and to a lesser degree the intergalactic medium (IGM). Subsequent generations of stars further process available gas reservoirs and increase the overall metal abundance, potentially creating a metallicity floor, or lower limit on the metal content in the IGM~\citep[e.g.][]{madau01}. Once sufficient enrichment levels are achieved in the CGM, metal signatures can be found in quasar absorption spectra, redward of the Ly$\alpha$ emission peak. Due to the brightness of the background quasar, absorption can be detected relatively easily from these diffuse gas phases even out to extremely high redshifts, whereas the detection of the associated low surface brightness emission would require more sensitive instrumentation and is currently restricted to lower $z$~\citep{bertone10a,bertone10b,bertone12,frank12,corlies16,corlies20,wijers22}.

At lower redshifts ($0 < z <1$), metals may be studied in tandem with nearby galaxies, especially those located within one to a few $R_\mathrm{vir}$. Such studies suggest possible correlations between galaxy traits and the covering fraction, strength, and incidence of metal absorption systems. These correlations also vary with the type of metal transition under consideration. Cool, low-ionization gas tracers like Mg II may show correlations in strength and incidence with the galaxy overdensity and also with the impact parameter, stellar mass, SFR, and specific SFR (sSFR) of the apparent host galaxy~\citep{chen10,bordoloi11,nielsen13,nielsen18,rubin18,fossati19,dutta20,dutta21,lundgren21,nelson21}. Warmer gas traced by for example C IV may increase in equivalent width and incidence with sSFR, and shows a flatter profile with impact parameter than cool gas tracers~\citep{oppenheimer08,borthakur13,ford13,bordoloi14,liang14,werk14,rubin15,dutta21,manuwal21}. C IV also may be less abundant and/or strong in the inner halo of massive galaxies~\citep[e.g.][]{burchett16}.

Progressing to higher redshifts it becomes increasingly difficult to associate any detectable galaxy with metal absorption systems. Searches performed with HST/ACS, MUSE, Keck-II, and ALMA have resulted in sporadic detections of galaxies, especially in Ly$\alpha$ emission, within the same field as known metal absorbers~\citep{diaz15,diaz21,cai17,wu21}. The distances between these objects vary widely, from $\sim$ 20 to hundreds of pkpc. These results have implications for the speed and mass loading factor of early galactic winds, the metal enrichment floor within the IGM, the sizes and luminosities of metal absorber host galaxies, and for determining the galaxies that initially contributed most of the metal budget. In some cases, the wind speeds necessary to provide the metals sourcing the detected absorption are far in excess of what is expected from simulations, and so the hosts are determined to be smaller, as yet undetected galaxies~\citep{garcia17}. Additionally, the small sizes of high-z galaxies and their accordingly shallow potential wells likely lead to low retention of metals, causing enrichment of a common group medium~\citep{porciani05}.

Coincident with the early enrichment process, the majority of gas in the Universe undergoes a phase transition at $z>5$, called the epoch of hydrogen reionization (EoR). The EoR lasts roughly from $6<z<12$, and is defined by the transition of the IGM from neutral to ionized as the ionizing photon budget increases. It is generally accepted to end at $z\sim6$ due to a rapid increase in scatter of Ly$\alpha$ opacity measurements at $z>5.5$, potentially indicating significant sightline-to-sightline variation in the hydrogen neutral fraction~\citep{fan06,becker15,bosman18,eilers18}. Although the obvious major influence of the EoR is on the ionization state of neutral hydrogen in the IGM, there is evidence that certain species of metal absorbers are also affected by the evolution in the UVB during this time. C IV incidence generally decreases towards higher redshifts~\citep[see e.g.][]{hasan20}, but especially rapid evolution is observed above $z=5$~\citep{codoreanu18,ryan-weber09,dodorico13,cooper19}. Similar trends, while less stark, also occur in Si IV~\citep{songaila01,codoreanu18,dodorico22}. Conversely, the low-ionization potential O I increases in abundance from $z>4.5$~\citep{becker19}. While decreasing enrichment towards higher redshifts can explain a decreasing abundance of metal systems, it alone cannot explain increasing incidence of O I. Further,~\citet{cooper19} find that a larger percentage of carbon-based absorption systems are found to include C II for redshifts greater than 5.7, suggesting evolution in the preferred ionization state of carbon.

Considering the evolution of metals at $z>5$ and their potential role as tracers of galaxies, it is useful to provide a better quantification of (1) the relationship between metal absorbers and galaxies near the end of the EoR, (2) the environments they occupy, and (3) what is driving their evolution during this time period. Towards this end, we have performed an investigation using the \emph{Technicolor Dawn} simulation, a cosmological hydrodynamic code including in-situ multifrequency radiative transfer, focusing on the period from $z=5$\textendash7. We build on previous studies by performing multifrequency radiative transfer to self-consistently model hydrogen reionization and the ionization states of metals. We also expand on our previous work with this simulation in~\citet{finlator20} by considering evolution with redshift and additional metal ions. In Section~\ref{sec:simulations}, we briefly summarize the relevant details of the simulations, describe our method for creating mock absorption spectra, and demonstrate our performance in reproducing several relevant observables. In Section~\ref{sec:results} we present our analysis of the metals' association with galaxies and environmental correlations. The results are placed in a wider context in Section~\ref{sec:discussion}, and we summarize the investigation in Section~\ref{sec:conclusions}.

\section{Simulations}\label{sec:simulations}
To investigate the association between metal absorption and galaxies after the end of reionization, we run cosmological hydrodynamic simulations and generate mock absorption spectra of metals and an associated catalog of detected systems. We assume a \emph{Planck} cosmology~\citep{planckcol16a}, with $\left(\Omega_M, \Omega_\Lambda, \Omega_b, H_0, X_H\right)=$ (0.3089, 0.6911, 0.0486, 67.74, 0.751). 

\subsection{Simulation Details}
We use the \emph{Technicolor Dawn} suite of simulations in this experiment, which simulates reionization using on-the-fly radiative transfer calculations~\citep{finlator18,finlator20}. The code is built on {\sc Gadget-3}~\citep[last described in][]{springel05}, a density-independent smoothed particle hydrodynamics (SPH) code, and uses methods from~\citet{hopkins13} to improve the treatment of fluid instabilities. The simulation box size is $l=15$ cMpc/h with 2 $\times$ 640$^3$ in baryons and dark matter, and radiative transfer is calculated in 80$^3$ voxels to define the local UVB. The RT implementation is multifrequency, with 24 frequencies spanning 1-10 Ryd to cover the ionization energies of a number of commonly observed metal transitions. Gas particles cool due to collisional excitation and ionization, recombination, dielectric recombination, and free-free emission~\citep{katz96}. Metal cooling is calculated on-the-fly under the assumption of collisional ionization equilibrium~\citep{sutherland93}.

Star formation occurs according to the subgrid~\citet{springel03} formulation, in which SPH particles with proper hydrogen number densities greater than $n_\mathrm{H}>0.13$ cm$^{-3}$ are capable of forming stars from a portion of their mass, at a rate appropriate for the Kennicutt-Schmidt law~\citep{kennicutt98}. A Kroupa~\citep{kroupa01} initial mass function (IMF) is assumed, with stellar masses ranging from 0.1-100 $M_\odot$. The ionizing emissivity is sourced by the star-forming gas particles, and determined using the spectral synthesis code {\sc Yggdrasil}~\citep{zackrisson11} according to the metallicity of the particles. Stars with masses $>$ 10 M$_\odot$ are immediately converted to SNe, with 50 per cent treated as hypernovae~\citep{nomoto06}; these then contribute heat and kinetic energy to galactic winds that drive metals into the CGM and IGM. We use a modified version of the stellar-mass dependent mass-loading factor formulation determined in~\citet{muratov15} from high-resolution {\sc Fire} simulations~\citep[see Eq. 1 of][]{finlator20}, as with use of this formulation without adjustment \emph{Technicolor} markedly underproduces the stellar mass function. Metal production by SNe is determined by weighting the yields of~\citet{nomoto06} to the Kroupa IMF. The contributions of asymptotic giant branch stellar winds are included as well, with yields from~\citet{oppenheimer09}. Through these pathways, the abundances of C, O, Mg, and Si (among others) are tracked for all particles. In post-processing, galaxies are identified using {\sc Skid}, and associated galaxy properties are calculated by interpolating the age and metallicity of star-forming particles according to FSPS~\citep{conroy10} to determine the spectral energy distributions, with additional accounting for line-of-sight effects such as redshifting and absorption by foreground H I.

\subsection{Metal Absorption Spectra}
In order to obtain our mock metal absorption spectra, we cast sightlines through the simulation volume, oblique to the boundaries, until a Hubble velocity of 10$^7$ km/s has been achieved. This allows the sightline to wrap around the simulation numerous times until it is well sampled. Gas particles encountering the sightline contribute to the metallicity, temperature, and peculiar velocity of the material giving rise to absorption, and the overall metallicity and metal mass fractions for several species are tracked for each SPH particle. The ionization states of each element are determined using the output ionizing background for the associated snapshot, which was calculated on-the-fly. This background includes a dominant spatially inhomogeneous contribution from the galaxy population and a sub-dominant spatially uniform quasar contribution, the latter of which assumes that quasars have formed but are rare and located far from the simulation volume. The ionization of metals is recalculated for the purpose of generating the spectra; both photo- and collisional ionization are taken into account, though collisional contributions are predicted to be sub-dominant.

The sightline is discretized into 2 km/s velocity segments, and the metal contributions to absorption are calculated assuming Voigt profiles~\citep{humlicek79}. Artificial Gaussian noise is added to each pixel for an S/N of 50, and metal absorption systems are identified where the flux drops $>$ 3 $\sigma$ below the continuum level for at least 3 consecutive pixels. The equivalent width (EW) of the systems is calculated, and converted to the column density that 
\begin{figure}
\includegraphics[width=0.44\textwidth]{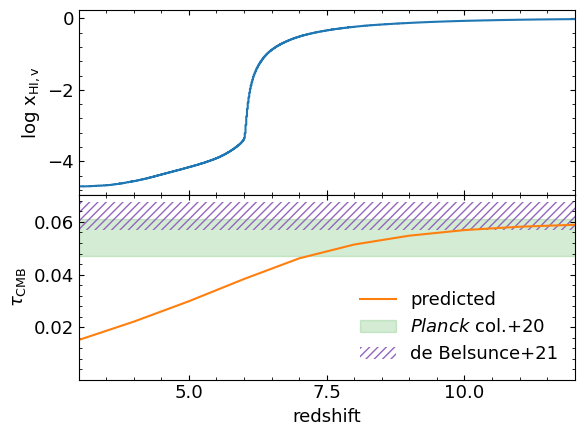}
\caption{The volume-weighted neutral hydrogen fraction (top panel) and the predicted $\tau_\mathrm{CMB}$ in \emph{Technicolor} from $z=12\rightarrow3$ (bottom panel). The~\citet{planckcol20} measurement is plotted as the green shaded region, while a re-analysis of the same data performed in ~\citet{debelsunce21} resulted in a slightly higher value lying at the center of the hatched region. The predicted simulation result lies within the upper 1$\sigma$ bound of the original \emph{Planck} analysis, and within the lower bound of the re-analysis.} \label{fig:predicted_tau_cmb}
\end{figure}
would be measured observationally using the apparent optical depth method~\citep{savage91}, assuming the sightlines are optically thin. An initial catalog is assembled of these systems and in keeping with observational practice, these are merged together within some threshold velocity separation to create the final catalog. For this we use a $\Delta v=50$ km/s~\citep[e.g.][]{dodorico13} and define the final centroid position as the column-weighted central wavelength, and the EW and column density $N$ as the sum of the individual contributing lines. The catalogs for all ions have no cutoff in column density or equivalent width, and range from $\sim$0.002 to 0.5\angstrom$\,$ (10$^{11.6}$ and 10$^{14.3}$ cm$^{-2}$). Where analysis is done for a subset of absorber strengths, it is noted in the text and/or figure caption.

\subsection{Verification of the simulation}\label{ssec:verification}
We briefly consider some basic metrics of the performance of the simulation against recent observational studies. For additional standard observational comparisons, see~\citet{finlator16,finlator18} and~\citet{finlator20}.

The predicted optical depth of the cosmic microwave background (CMB) photons to Thomson scattering can be calculated as
\begin{equation}\label{eq:thomson_scattering_optical_depth}
    \tau_\mathrm{CMB} = \int_{0}^{z_\mathrm{ls}} \frac{c\;\sigma_\mathrm{T} \; n_e\left(z\right)}{\left(1+z\right) \; H\left(z\right)} dz 
\end{equation}
where $\sigma_\mathrm{T}$ is the cross-section for Thomson scattering, $n_\mathrm{e}$ is the number density of electrons, $H\left(z\right)$ is the Hubble parameter, and $z_\mathrm{ls}$ is the redshift at the surface of last scattering. $n_\mathrm{e}(z)$ is calculated from the global volume-weighted neutral fraction. $x_\mathrm{HeI}$ and $x_\mathrm{HeII}$ are assumed to follow $x_\mathrm{HI}$ and $x_\mathrm{HII}$ until helium reionization occurs at $z=3$, at which point all helium transitions to the He III state. Including these contributions from H II and singly and doubly ionized helium, we find $\tau_\mathrm{CMB} = 0.0615$ (Figure~\ref{fig:predicted_tau_cmb}), compared to the Planck measurements of 0.054 $\pm$ 0.007~\citep{planckcol20}. A re-analysis of the Planck data using different formulations of the likelihood arrived at a similar but slightly higher $\tau_\mathrm{CMB}=0.0627^{+0.0050}_{-0.0058}$~\citep{debelsunce21}. Thus, our reionization history is consistent with the observations, even when accounting for different interpretations of the data.

The Ly$\alpha$ transmission is a function of the optical depth to Ly$\alpha$ photons, and calculated as $T_\mathrm{Ly\alpha} = e^{-\tau}$ (Figure~\ref{fig:SIM_figure_2}). To predict the optical depth in Ly$\alpha$ for the simulation output, we use the same mock sightline as is used to generate simulated metal absorbers. Since the sightline is $>$ 10 Gpc in length for all redshifts and wraps through the entire volume, this should accurately reflect the global simulated $T_\mathrm{Ly\alpha}$ of the simulation box, although due to the small box size this could deviate from the true value because of cosmic variance. We find that the mean transmission is in agreement with the broad trends implied by QSO observations, although it deviates beyond 1 $\sigma$ for several of the data points, particularly for $z<5.4$ for~\citet{bosman21}.

Given that Lyman Alpha Emitters (LAEs) are a common target of high $z$ galaxy searches, it is of interest to model the Ly$\alpha$ luminosities of our galaxy population. To predict the baseline Ly$\alpha$ luminosities of our simulated galaxies, we use the empirically derived relation from~\citet{oyarzun17} relating Ly$\alpha$ equivalent widths to the stellar mass\footnote{In reality, predicting the detected Ly$\alpha$ luminosity of galaxies is complex due to the resonance of the transition and the physical configuration of the gas in the ISM, CGM, and IGM, which may motivate a more sophisticated method of estimation. Further, the empirical relation in Equation~\ref{eq:mstar_to_LLya_EW} is measured at $z=3$\textendash4, where IGM attenuation is not as significant as for $z>5$. However, we use this method to maintain consistency with ~\citet{finlator20}}:
\begin{figure}
\includegraphics[width=0.44\textwidth]{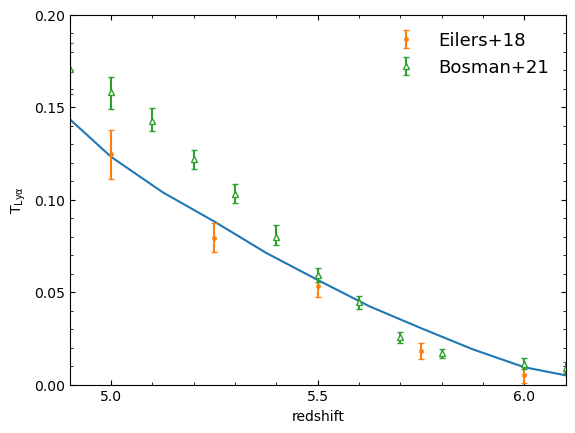}
\caption{The mean Ly$\alpha$ transmission calculated from the mock sightline, overplotted with the measurements from QSO spectra~\citep{eilers18,bosman21}. The observational results generally match the trend produced in the simulation, although the evolution may be too shallow, especially compared against~\citet{bosman21}.} \label{fig:SIM_figure_2}
\end{figure}
\begin{equation}\label{eq:mstar_to_LLya_EW}
    \log_\mathrm{10} \mathrm{EW} = \log_\mathrm{10} M_* + 190
\end{equation} 
From this relationship, we predict the Ly$\alpha$ luminosity function (LALF) in Figure~\ref{fig:lalf_and_smf} (top panel), further overplotting observational 
\begin{figure}
\includegraphics[width=0.44\textwidth]{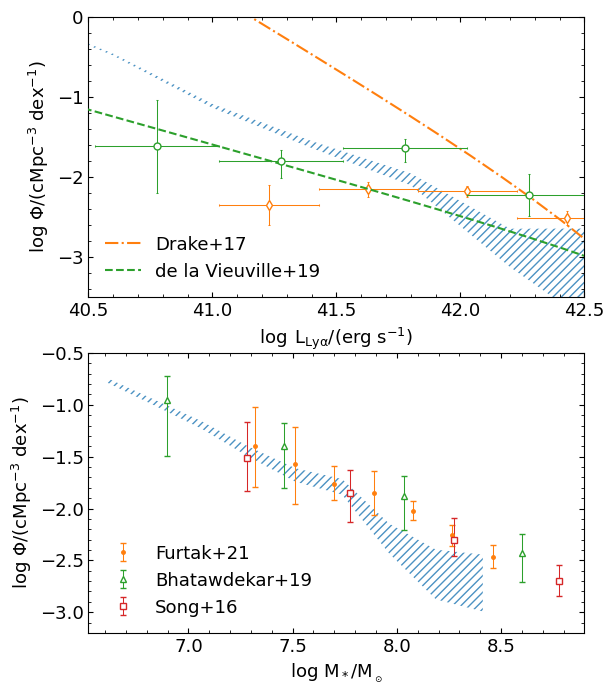}
\caption{Top panel: The fiducial $z=6$ Ly$\alpha$ luminosity function from the simulation (blue hatching) obtained using Equation~\ref{eq:mstar_to_LLya_EW} to estimate $L_\mathrm{Ly\alpha}$. $V_\mathrm{max}$-corrected observations from~\citet{delavieuville19} and their best fit relation are shown as green points and the dashed line, respectively. Uncorrected datapoints from~\citet{drake17} are shown in orange while their completeness-corrected maximum-likelihood relation is the dash-dot orange line. The observational methods of completeness correction result in widely varying LALFs, with our fiducial relation lying in between. Bottom panel: The $z=6$ stellar mass function with 1$\sigma$ scatter overplotted with several observations and best-fit relations~\citep{song16,bhatawdekar19,furtak21}. For $\log M_*/M_\odot < 7.8$ the simulation matches the observations results, while for higher stellar masses the number of galaxies falls short.} \label{fig:lalf_and_smf}
\end{figure}
results from $z=5-6.7$ that probe to low luminosities. Our predictions lie within the extremes of the completeness-corrected Schechter function fits from~\citet{drake17} and~\citet{delavieuville19} for $\log L_\mathrm{Ly\alpha}$/(erg s$^{-1}$)$<$42. Although we exclude them from this figure, LALFs at $z\sim6$ are observed to extend to significantly higher luminosities~\citep[e.g.][]{santos16,konno18}. While a general underproduction of higher Ly$\alpha$ galaxies could result from our $L_\mathrm{Ly\alpha}$ prescription, the precise redshift under consideration, or the simulation box size, in this case the complete cut off at $10^{42.5}$ erg/s certainly results from our small box size being unable to sample rarer, brighter galaxies. Although our predictions for the most part lie within the bounds of the observations, due to differences in these works' methods of completeness correction, there is strong disagreement in the predicted underlying populations of Ly$\alpha$ emitting galaxies, making it difficult to determine the overall accuracy of our predicted population.

Given the large uncertainty on the observed LALF, and the potentially large variation between the predicted LALF and observations, we also examine the predicted stellar mass function (SMF) against recent observations to determine whether the model accurately reproduces this metric (Figure~\ref{fig:lalf_and_smf}, bottom panel). We find that for stellar masses below 10$^{7.8}$ M$_\odot$, our predictions lie within the uncertainties from several studies probing to low stellar masses at $z\sim6$, although they tend to lie systematically low compared to the best-fit Schechter relations. At masses $M_*>$ 10$^{7.8}$ M$_\odot$ the simulation underproduces the incidence of galaxies, likely due in part to the small volume used in this study. This has implications for our predictions of galaxy detections near metal absorption systems, since this regime of higher stellar masses will be brighter in Ly$\alpha$ as well.

\begin{figure}
\hspace*{-0.2in}
\includegraphics[width=0.51\textwidth]{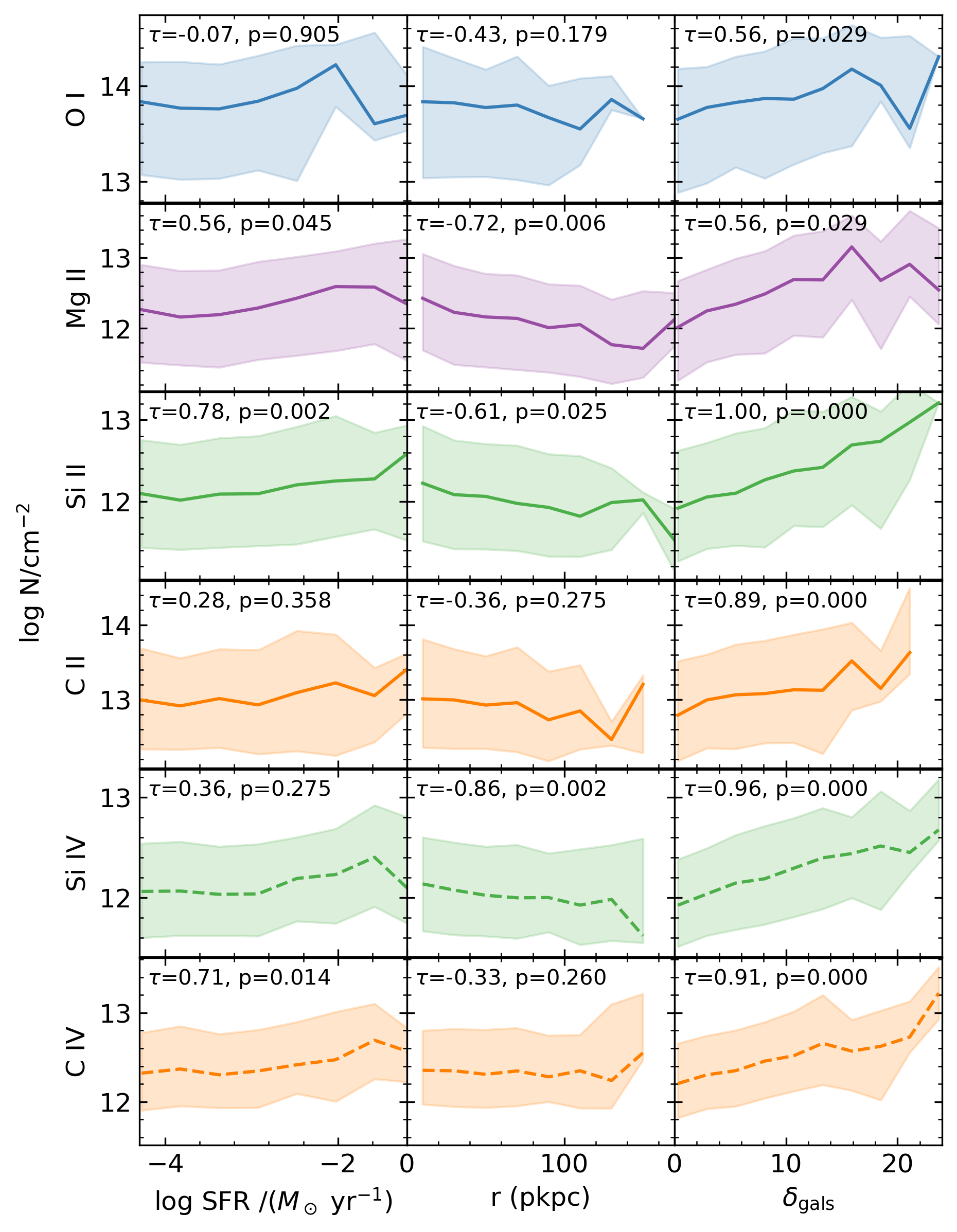}
\caption{The absorber column density $\log N$/cm$^{-2}$ as a function of the log SFR and impact parameter of the nearest galaxy, and the local overdensity within 100 pkpc at $z=6$. Note that the y-axis limits are different for each ion. The lines indicate the median log $N$/cm$^{-2}$ in each bin, while the shaded regions indicate the 16th-84th percentile spread of the distribution, and the Kendall $\tau$ and associated p-value are noted in the upper left of each panel. For the SFR column $\tau$ is calculated on the unlogged SFR. While each relationship shows considerable scatter, there are trends of increasing absorber strength with host log SFR and local overdensity, and decreasing strength with impact parameter. The general trends in these parameters are the same at $z=5$ and $z=7$, only varying slightly in their correlation coefficients.} \label{fig:trends_with_logN}
\end{figure}
\section{Results}\label{sec:results}
In this section we describe the model predictions for the number and observability of galaxies around metal absorbers from $z=5$\textendash7, the relationships between metal absorbers and their host galaxy traits, and the characteristics of the environments probed by the metals. This is done with an eye towards determining the ideal candidates for future surveys targeting high redshift galaxies.
\begin{figure*}
\includegraphics[width=0.85\textwidth]{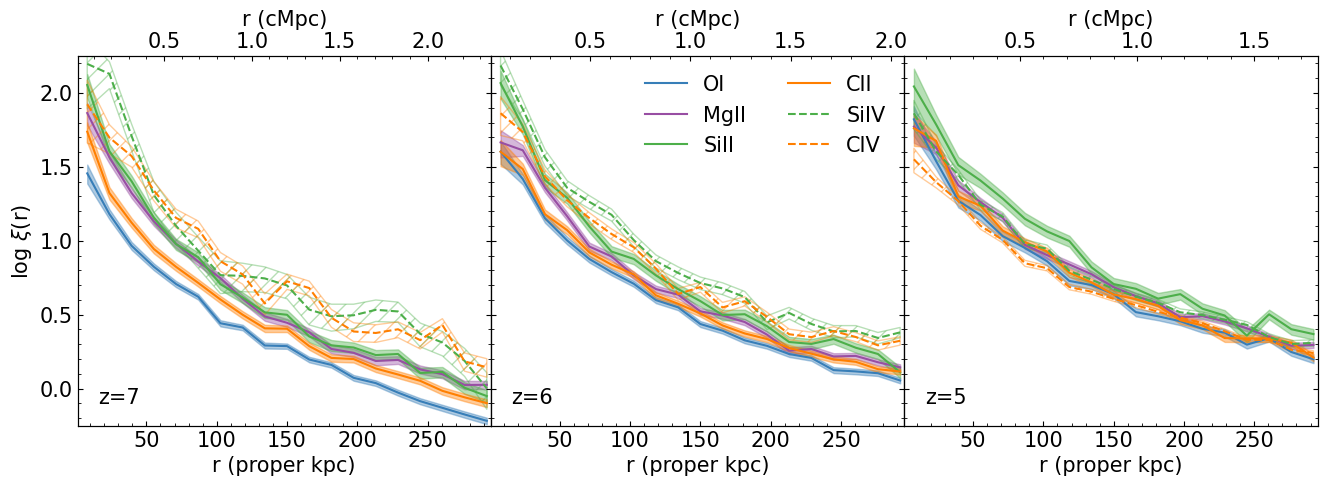}
\caption{The galaxy-absorber cross-correlation functions for each transition at three redshifts spanning the end of reionization, with galaxies restricted to those with $\log L_\mathrm{Ly\alpha}$/(erg s$^{-1}$)$>40.5$, and absorption systems restricted to $>$10$^{13}$ cm$^{-2}$. The error is calculated by assuming a Poisson-like error on the counts, $\sqrt{\mathcal{N}}$, and propagating this into $\log \xi_\mathrm{g-abs}$. The cross-correlations for Si IV and C IV are elevated with respect to those of the lower ionization transitions at $z=7$ and 6, but this dissipates by $z=5$. Metals are associated with an excess of galaxies out to at least 200 pkpc by $z=7$, and the higher ionization metals out to 300 pkpc. The distance out to which galaxies and absorbers are positively correlated increases for all ions at lower redshifts.} \label{fig:OAT_figure_2}
\end{figure*}

\subsection{Observability and Traits of Nearby Galaxies\label{ssec:OAT}}
For $z<2$, correlations are observed between metal absorbers and the characteristics of their host galaxies; thus we begin by looking for any such correlations in our simulated systems. At $z=6$ we first examine the relationships between the column density of systems and the SFR of the galaxy that is closest in real, rather than projected, distance (Figure~\ref{fig:trends_with_logN}). Absorber positions are identified according to their column-density-weighted velocity centroid, in effect similar to the manner in which they would be interpreted observationally. This neglects the effects of redshift space distortions due to the peculiar velocity of the gas, which may significantly affect the apparent physical location of the absorbing gas~\citep[see e.g.][]{peeples19,marra22}. These closest galaxies are taken to be the hosts of the metal absorbers, regardless of any true physical association.

The absorber strengths generally show an increase with the star formation rate, at least for $\log \mathrm{SFR}$/($M_\odot$ yr$^{-1}$)$>$ -3, although there is considerable scatter in the relation. Absorber strengths as a function of impact parameter from the host galaxy are shown in the second column of Figure~\ref{fig:trends_with_logN}. Here we find the same general trend as seen in observations, namely that the typical absorber strength decreases as one probes distances farther from the host. We also examine the absorber strengths with respect to galaxy overdensity, defined as $\delta_\mathrm{gals} = \rho_\mathrm{gals}/\langle \rho_\mathrm{gals}\rangle - 1$, where $\rho_\mathrm{gals}$ is the number density of galaxies within 100 pkpc. We find a correlation between the overdensity and the absorbing system strength, indicating that stronger absorption systems should be probing larger galaxy overdensities on average. The typical overdensities for different transitions show more variability than did the SFR or distance, for example showing a 50 per cent greater local overdensity around Si IV absorbers compared to O I systems at $z=6$. 

Calculating the Kendall $\tau$ rank correlation coefficient for bins of each ion's strength and the \emph{unlogged} SFR, we find that Mg II, Si II, and C IV have statistically significant correlations with an average correlation coefficient of 0.68. Between the overdensity and log N, we find statistically significant correlations for all ions with an average of 0.81. In cases where an ion's strength is correlated with both the SFR and overdensity, the association is usually stronger with the overdensity. Of course, there is some intermingling between these metrics, as the average SFR of galaxies in denser environments tends to be higher~\citep{koyama13,cooke2014,shimakawa17,ito20}. It also seems likely that there would be a stronger, less sporadic trend with host SFR were the simulation volume larger, therefore allowing for a larger dynamic range in SFR and overdensity and the formation of galaxies with higher SFRs, as the maximum SFR here only reaches 0.12 $M_\odot$ yr$^{-1}$.

The physical separations between the absorber and the nearest galaxy span a wide range, from $<$ 1 to over 100 pkpc, so it is difficult to interpret the extent to which the absorbers are truly physically associated with their nearest galaxy. Future runs of \emph{Technicolor} will incorporate particle tracing, and therefore allow subsequent investigations to determine the true sources of the metal systems. These results may also be affected by the tendency of the simulation to underpredict (overpredict) strong (weak) absorption systems, possibly arising from a general tendency in simulations to eject metals too efficiently from small galaxies, which we discuss more in Section~\ref{sec:discussion}. Intuitively, the over-ejection of metals would result in a generally larger physical separation between the metal systems and their sourcing galaxy.

To predict detections of galaxies near metal absorption systems, we first calculate the galaxy-absorber cross-correlation function, or the average overdensity of galaxies around absorbers as a function of distance:

\begin{equation}\label{eq:overdensity_xi}
\xi_\mathrm{gal-abs} \left( r \right) = \frac{1}{n_0}\frac{\bar{N}(r)}{\Delta V} - 1
\end{equation}
where $n_0$ is the global number density of galaxies with $L_\mathrm{Ly\alpha}$ greater than a given luminosity, $\bar{N}\left(r\right)$ is the average number of such galaxies around a given metal absorber, and $\Delta V$ is the volume contained within a spherical shell of radius $r$. We plot this function for each metal type for $z=5$, 6, and 7 in Figure~\ref{fig:OAT_figure_2}, with errors assumed to be Poissonian. All ions are associated with an excess of galaxies out to at least 200 proper kpc at $z=7$, and the association naturally extends to larger physical distances at lower redshifts, as gas is enriched out to lower overdensities by galactic winds. The cross-correlations with galaxies of high ionization ions Si IV and C IV are higher than those of the low ionization ions out to 300 pkpc at $z=7$, and this is still generally true by $z=6$, although the deviation is smaller. By $z=5$ there is no longer an obvious distinction between the curves for high and low ionization ions.

\begin{figure*}
\includegraphics[width=0.8\textwidth]{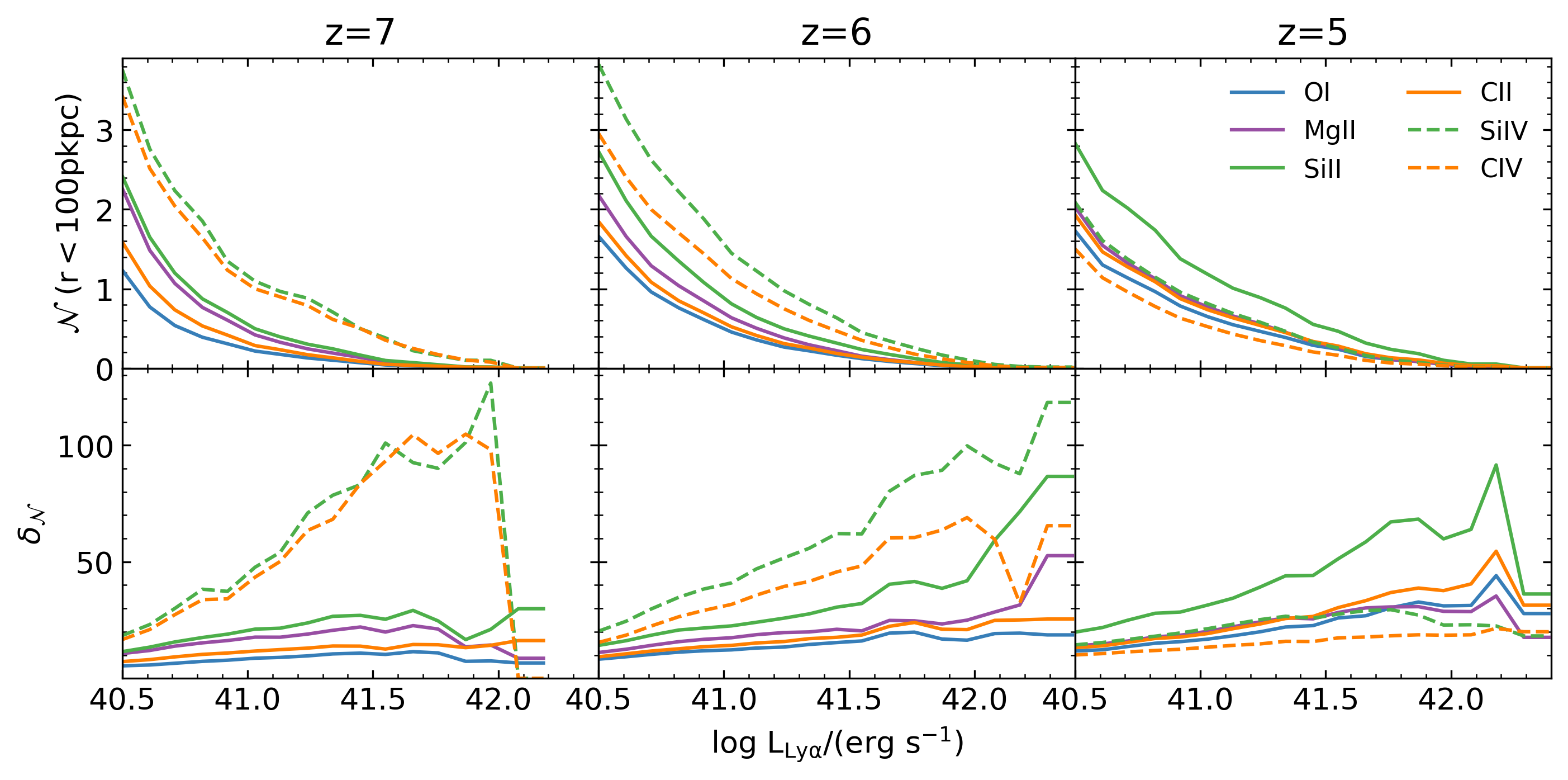}
\caption{Top row: The predicted number of galaxies around each absorber type as described in Equation \ref{eq:galaxy_count}, as a function of the galaxy Ly$\alpha$ luminosity, and for three integer redshifts from $z=7\rightarrow5$. Bottom panel: The excess of the expected galaxy counts with respect to the number of field galaxies with the given luminosity. For redshifts prior to the end of reionization at $z\sim6$, high ionization metal transitions are associated with a larger number of galaxies for nearly all considered brightnesses. By $z=5$, this trend has diminished.}\label{fig:OAT_figure_3}
\end{figure*}
To determine the average total number of galaxies within 100 pkpc of an absorber, we integrate over Equation~\ref{eq:overdensity_xi}:
\begin{equation}\label{eq:galaxy_count}
\mathcal{N} = \int_0^{100\; \mathrm{pkpc}} 4 \pi r^2 n_0 \left( 1+\xi_\mathrm{gal-abs}\right) dr
\end{equation}
In Figure~\ref{fig:OAT_figure_3} we show the calculated $\mathcal{N}$ around metal systems with column density log $N$/cm$^{-2}>$13, including redshift evolution from $z=7\rightarrow5$ in the different columns. We find that while the average number of galaxies within 100 pkpc around differing absorber types are on the order of a few and relatively similar to one another, at $z=7$ and $z=6$ there is a greater average incidence of galaxies near high-ionization ions C IV and Si IV, which follows logically from their greater excess shown in Figure~\ref{fig:OAT_figure_2}. By $z=5$, this trend disappears, and most ions have roughly equal numbers of nearby galaxies, with the exception of a prominent increase around Si II. In fact, at $z=5$ C IV has the fewest predicted number of nearby galaxies.

We also show the overdensity as a function of Ly$\alpha$ luminosity and plot these in the lower panels of Figure~\ref{fig:OAT_figure_3}. The excess shows that the deviation from the expected numbers is greater around galaxies that are brighter in Ly$\alpha$, particularly at higher redshifts. By $z=5$, the converging excesses highlight the more similar environments traced by most transitions, although they are not exactly the same. From these fiducial results, we would predict that all ions with log N/cm$^{-2}$>13 would have 1 or more galaxies with $\log L_\mathrm{Ly\alpha}$/(erg s$^{-1}$)$\geq 40.5$ within 100 pkpc at each of these redshifts, but that for galaxies one dex brighter in Ly$\alpha$, less than one will be present on average.

The observational completeness for galaxies with L$_\mathrm{Ly\alpha}=10^{42}$ erg/s is only on the order of 50 per cent, and the estimates for the abundances of fainter galaxies vary widely due to this uncertainty. Regardless, for log Ly$\alpha$/(erg s$^{-1}$)$\sim40.5$, completeness should be extremely low. While the raw numbers of nearby galaxies we predict are not large, the excess measurement still indicates a greatly elevated likelihood of detecting galaxies near metal absorbers, and odds that vary depending on the ion and the redshift.

To examine the effect of our $L_\mathrm{Ly\alpha}$ models on predicted detections in Section~\ref{sec:results}, we adjust the predicted luminosities by abundance matching our galaxy catalog to the observed luminosity functions, retaining the overall number of galaxies and the \emph{in situ} galaxy-absorber relationship while adjusting only the predicted Ly$\alpha$ luminosity. We perform two adjustments to match the predictions to the relatively high incidence in~\citet{drake17} and the low incidence in~\citet{delavieuville19}.

\begin{figure}
\includegraphics[width=0.44\textwidth]{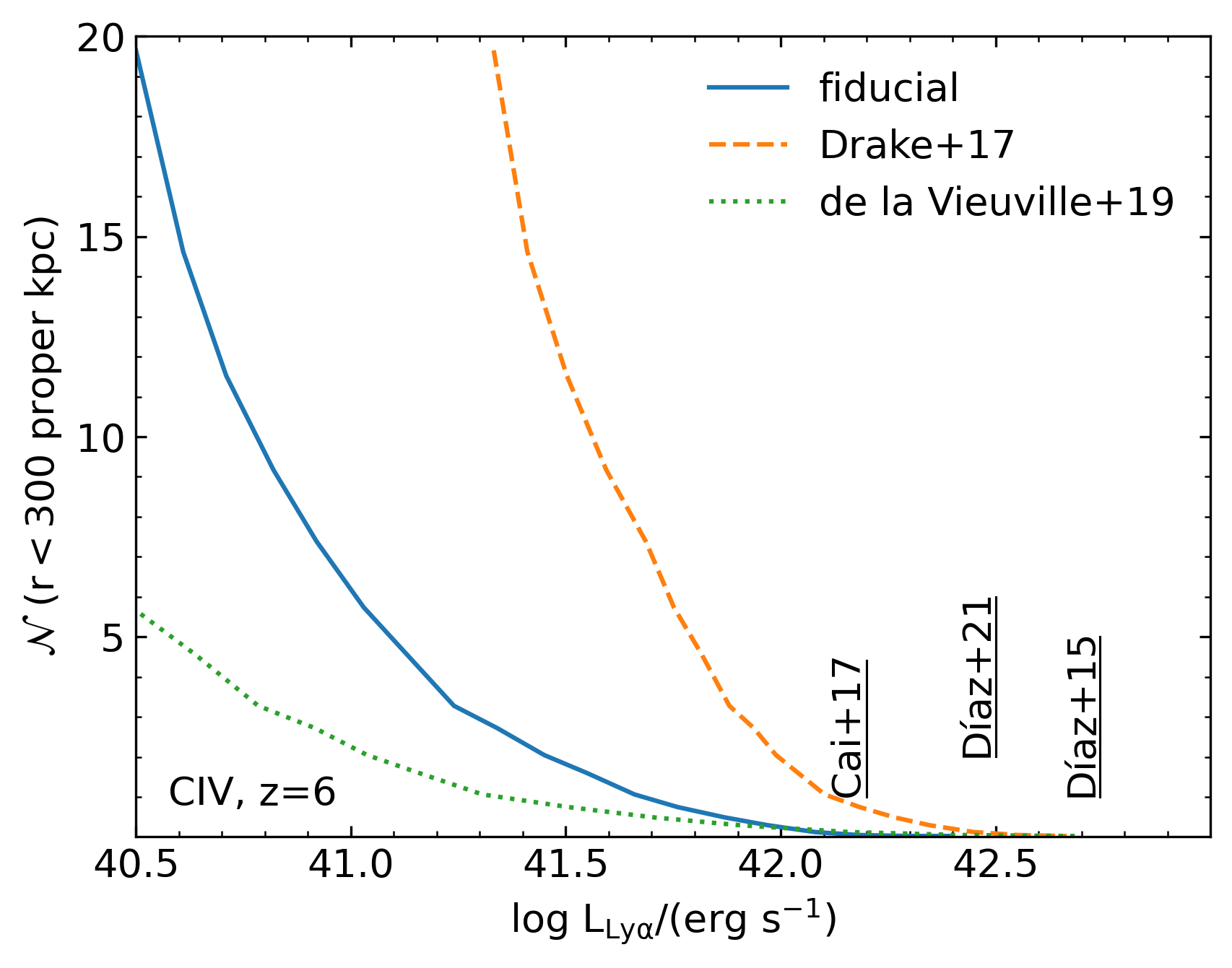}
\caption{The predicted number of galaxies within 300 pkpc of C IV systems based on the results from our $z=6$ snapshot, one with the fiducial $M_*$-L$_{Ly\alpha}$ relationship and two where it has been adjusted. This comparison represents a scenario in which the galaxy-metal relationship is properly recreated in the simulation, but where the Ly$\alpha$ luminosities of the galaxies are not properly estimated, i.e. if the~\citet{oyarzun17} formulation is not truly applicable for these redshifts or galaxy masses. The blue line indicates the fiducial relationship, while the orange and green lines show the predicted number of galaxies when adjusted for the~\citet{drake17} and~\citet{delavieuville19} LALFs. Overplotted are several observations of LAEs near known C IV systems from $z>5$, where the base of the annotation indicates the number of detected galaxies and the horizontal position indicates the average luminosity. Assuming the veracity of the SMF and the galaxy population, these results indicate either a mis-calculation of the Ly$\alpha$ luminosity or an overproduction of metals near faint galaxies.} \label{fig:OAT_figure_4}
\end{figure}

Abundance matching our galaxy catalogs to either~\citet{delavieuville19} and~\citet{drake17}, and expanding the search area to 300 pkpc, we predict the number of detections around C IV systems with $\log$ $N>13.0$/cm$^{-2}$ in Figure~\ref{fig:OAT_figure_4}.\footnote{We focus on C IV here because it has been a common target for such high-z galaxy searches, and thus there are several observational data points available for comparison.} For $\log$ $L_\mathrm{Ly\alpha}$/(erg s$^{-1}$)$>$41.5, the~\citet{delavieuville19} LALF would result in about one galaxy detection within 300 pkpc, whereas the fiducial simulation predicts about two and~\citet{drake17} about 12. For a more plausibly detectable $\log$ $L_\mathrm{Ly\alpha}$/(erg s$^{-1}$)$=42$ galaxy, the~\citet{drake17} LALF results in about one LAE near C IV absorbers, while the other two models have less than one on average. For comparison, we overplot detections of LAEs near C IV systems at $z\sim5.7$. In combination, these results predict 1-2 detections of LAEs with $\log$ $L_\mathrm{Ly\alpha}$/(erg s$^{-1}$)$>42.2$ which lies higher than any in our fiducial model.

Physically, Ly$\alpha$ emission from galaxies is dictated by the intrinsic emission generated within the ISM and the Ly$\alpha$ photons' interaction with the surrounding CGM and IGM, so it is difficult to know if our Ly$\alpha$ model is reasonable. The ISM alone is structurally complex and unresolved in cosmological simulations~\citep[e.g.][]{smith22}, and thus the relationship between the escaping signal and galaxy characteristics is uncertain. Adding to the complexity, the relation in~\citet{oyarzun17} is assuredly less affected by attenuation than any LAEs observed at $z\geq5$ due to the increasing gas density and overall higher neutral hydrogen fraction at high $z$. However, the same general trends we report in log L$_\mathrm{Ly\alpha}$ are also apparent in the trends of galaxy UV magnitude, with the numbers of bright galaxies being enhanced near high ion absorbers prior to $z=6$ and diminishing towards lower redshifts. For example, at $z=6$ where the JWST blank-field limit is $M_\mathrm{UV}=-15$, there are on average 9 galaxies within 300 kpc of high ion absorbers and 6 near low ions, and the differences are more stark at higher $z$.

\subsection{Environments of Absorbers}\label{ssec:ENV}
\begin{figure*}
\includegraphics[width=0.8\textwidth]{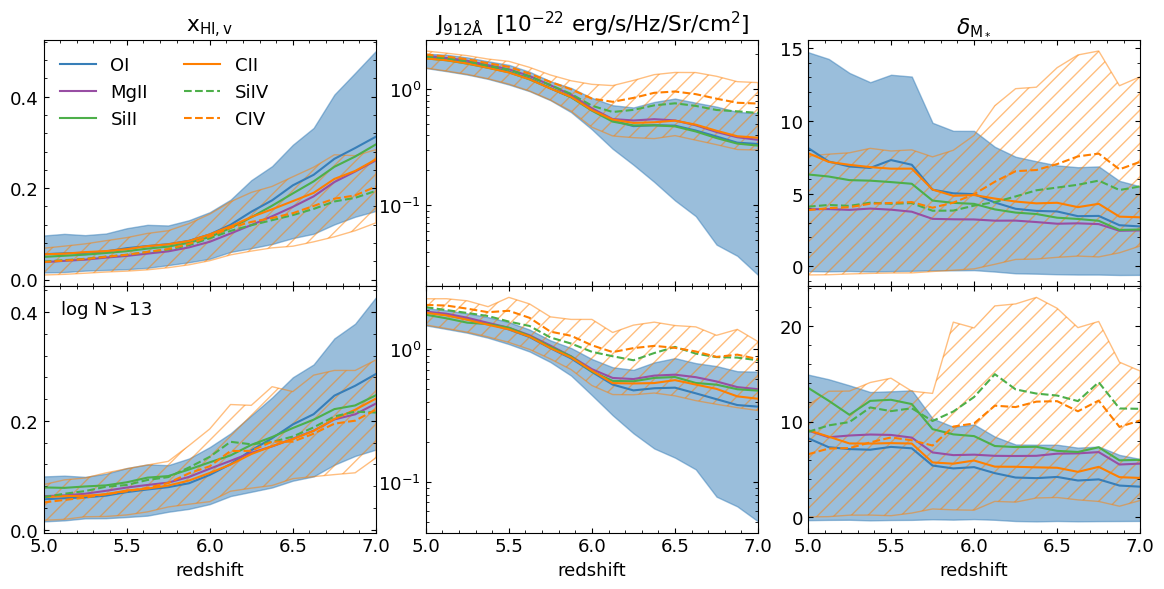}
\caption{Top row: Evolution with redshift of the mean specific intensity at $\nu=912\angstrom$, H I neutral fraction, and stellar mass overdensity for all metal systems regardless of strength, for the environment within $\sim$ 100 pkpc. The line styles and colors vary with the ion type, and shaded and hatched regions show the 16th-84th percentile spread for O I (blue) and C IV (orange). Bottom row: The same redshift evolution, but for metal systems with log N/cm$^{-2}>$13. Prior to the end of reionization at $z\sim6$, C IV and Si IV trace regions with higher $J_\mathrm{912}$ and slightly higher stellar mass overdensities.} \label{fig:ENV_figure_1}
\end{figure*}

The greater numbers of galaxies predicted around high-ionization absorbers at $z\geq6$ demonstrate that they trace higher overdensities of galaxies at these times, especially those brighter in Ly$\alpha$, than do low-ionization metals. To isolate the cause of this difference and determine the drivers of their environmental preferences, we directly examine the local specific intensities, neutral fractions, and stellar mass overdensities in an $r=100$ pkpc sphere around these ions within the simulation and how they evolve with redshift (Figure~\ref{fig:ENV_figure_1}). We show the $1\sigma$ deviation around O I and C IV as example cases of low and high ion trends. For the weak absorbers (top row), in keeping with our expectations from Section~\ref{ssec:OAT}, we find that the high ionization ions C IV and Si IV initially have a higher mean stellar mass overdensity $\delta_\mathrm{\rho_*}$($=\rho_*/\langle \rho_* \rangle-1$) than the other ions, but this average decreases with decreasing redshift. For example, $\delta_\mathrm{M_*}$ around C IV decreases from 7.2 at $z=7$ to 3.9 at $z=5$. On the other hand, the low ionization states all show increases in terms of their average overdensities over the same period, with O I for example increasing from 2.7 to 8.1.

The association of high-ionization absorbers with higher galaxy stellar mass overdensities could indicate that the physical conditions in these regions are preferable for either their formation and/or detection. In particular, they could in theory be arising due to a locally amplified ionizing background or a locally reduced neutral hydrogen fraction, ($x_\mathrm{HI}$) and thus reduced shielding from an ionizing background. Additionally, it could result from their relative rarity, and thus a tendency to arise in regions with higher gas overdensities and where greater quantities of metals would be expected.

We find that the mean neutral fractions are quite comparable between all ions by $z=7$, with O I and C IV for example with 30 and 20 per cent, respectively. The main difference between ions lies in their 16th-84th percentile distributions for a given redshift, with O I ranging from 15-50 per cent and C IV from only 12-28 per cent neutral, showing a tendency for low ionization states to probe a tail of unusually neutral gas. With decreasing redshift, this general trend continues to describe the distributions of the low and high ionization ions, although with the median $x_\mathrm{HI}$ gradually decreasing for both groups.

The specific intensity at the Lyman limit, $J_\mathrm{912\angstrom}$, is strongly related to the hydrogen neutral fraction; in regions where $J_\mathrm{912\angstrom}$ is high, the neutral fraction generally will be low. Accordingly, we find that the low ionization ions that probe unusually neutral gas at $z>6$ also have a tendency to arise in gas with a low specific intensity, with the mean value $\sim$2 times higher for C IV than for O I at $z=7$. The distributions gradually converge as they approach $z=6$, and the medians and ranges are more similar by $z=5$. 

Restricting the metal systems to only those with $\log N$/cm$^{-2}>13$ affects the details of these trends. In particular, at $z=7$ for $x_\mathrm{HI}$ the high and low ions have more similar average values although the 16th percentiles of the high ions' distributions still skew lower than for low ionization ions, and the distributions quickly converge at lower z. The UVB amplitude on the other hand remains elevated around stronger high ions down to $z=5$, in contrast to the converging means for weak high and low ions. The discrepancy in stellar mass overdensities between low and high ions at $z=7$ is also greater for the stronger absorbers, although they converge by $z=5$.

The tendency of high ionization systems above $z=6$ to trace regions which are slightly more ionized, more overdense with stars, and that have higher Lyman limit specific intensities shows a preference for regions with higher stellar mass galaxies. The decreasing association of high ions with galaxies below $z=6$ could make sense if greater stellar masses led to increased enrichment and thus higher metal column densities at larger radii, in turn increasing the number of detectable metal systems at large impact parameters. However, if this were the case then this migration away from higher overdensities should also be seen in the low ions, while in fact the reverse is true.

\begin{figure*}
\includegraphics[width=0.8\textwidth]{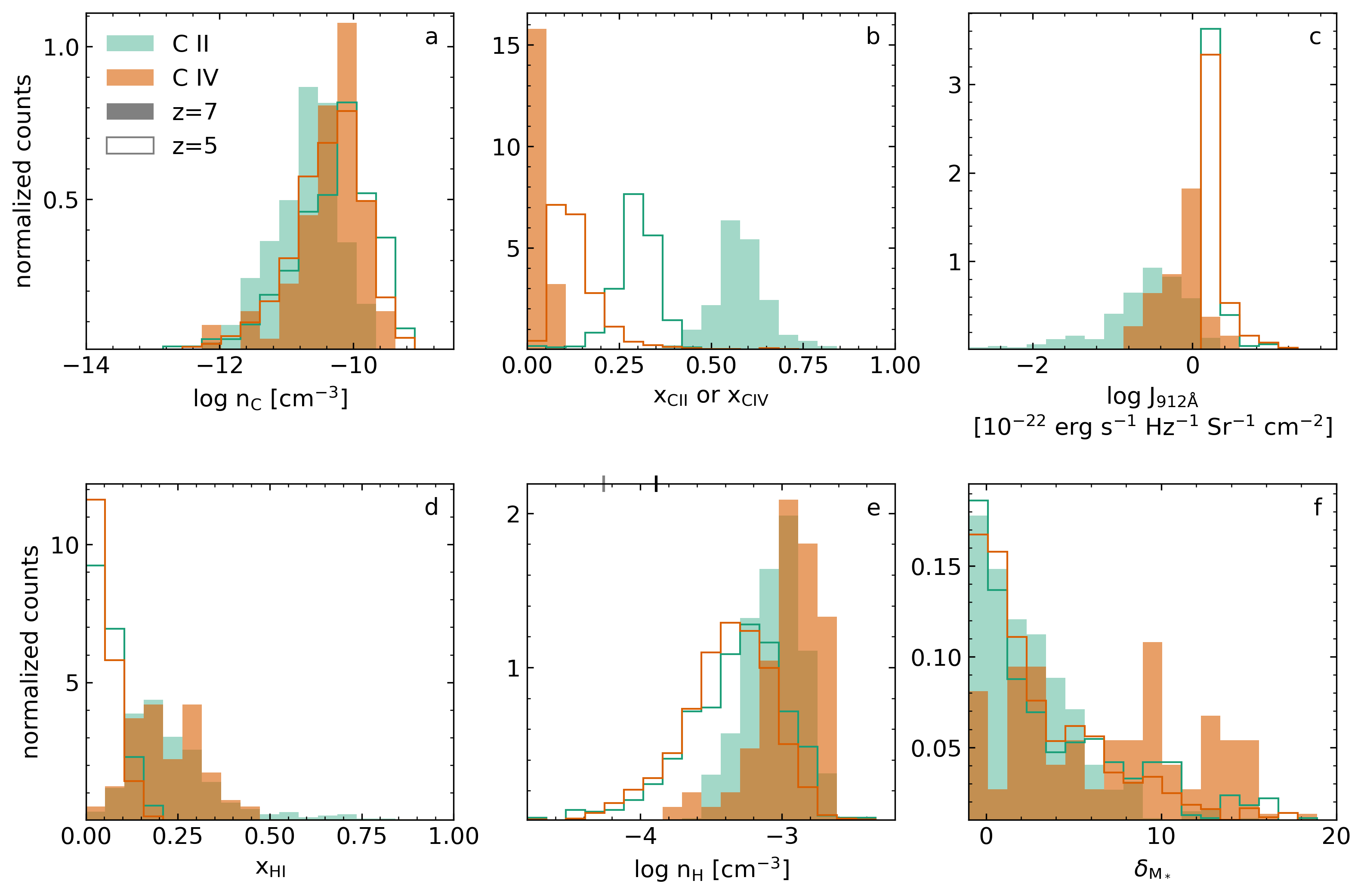}
\caption{The physical properties of $l=100$ pkpc cells containing $\log N$/cm$^{-2}>13$  C II and C IV systems (teal and ochre, respectively) at $z=7$ and $z=5$ (filled and empty). The mean density at $z=7$ and $z=5$ are shown as ticks (black and gray, respectively) on the upper axis in panel e. The means of the distributions are shown in Table~\ref{tab:property_averages}. The ionization fractions in panel b show the mass fraction in the ion of interest, for example the solid ochre shows the mass fraction in C IV of C IV-absorbing cells at $z=7$. Strong carbon-absorbers usually arise in gas that is denser, more enriched, and located more closely to galaxies. Since reionization concluded at $z=6$ in the simulation, the environments capable of hosting high-ionization C IV have increased, causing C II and C IV to probe increasingly similar regions, although C II is associated with higher stellar mass overdensities and slightly higher gas densities.}\label{fig:ENV_figure_2}
\end{figure*}

To investigate the opposing trends seen in low and high ions, we examine the histograms of the simulation cell properties for gas containing strong C IV and C II systems for both $z=7$ and for $z=5$ (Figure ~\ref{fig:ENV_figure_2}), and compare them to the global gas properties given in Table~\ref{tab:property_averages}. We consider the number density of carbon, the mass fraction of C II and C IV, the Lyman limit specific intensity, neutral fraction, gas density, and stellar mass overdensity. \footnote{The values in cells around each absorber are averaged, and given the relatively large search area, result in values being skewed somewhat towards the global average properties. This does not necessarily indicate, for example, that the majority of the metal absorbers are arising in diffuse IGM gas.} First, at $z=7$, the carbon-absorbing gas (colored shaded regions) is generally more dense, carbon-rich, and has a higher local average stellar mass overdensity than the global cell properties in the simulation. This carbon-absorbing gas also has a greater local UVB amplitude at the Lyman limit and is more ionized with a lower H I neutral fraction.

\begin{table}
  \tablewidth{\columnwidth}
  \caption{Mean properties of the distributions in Figure~\ref{fig:ENV_figure_2} of C II- and C IV-absorbing gas cells. Global means are included for comparison. The global quantities are calculated by summing total quantities and dividing by the total simulation volume, while the quantities around absorbers are averages of the surrounding cells within 100 pkpc. The choice in the size of the cell affects the range of possible values; for example, calculating this with smaller cells would result in a wider range of possible densities and overdensities.\label{tab:property_averages}}
  \begin{tabular}{ c  r r r r r r}
    \hline
    \hline
    \multirow{2}{*}{} &
      \multicolumn{2}{c}{Global} &
      \multicolumn{2}{c}{C II} &
      \multicolumn{2}{c}{C IV} \\
    & $z=7$ & $z=5$ & $z=7$ & $z=5$ & $z=7$ & $z=5$ \\
    \hline
    log $n_\mathrm{C}$   & $-$11.2 & $-$10.98 & $-$10.64 & $-$10.45 & $-$10.32 & $-$10.54 \\
    $x_\mathrm{CII}$     & 0.57    & 0.29     & 0.58     & 0.27      &  0.53    & 0.25     \\
    $x_\mathrm{CIV}$     & 0.06    & 0.11     & 0.02     & 0.15      & 0.04     & 0.17  \\
    log J$_\mathrm{912}$ & $-$0.67 & 0.28     & $-$0.61  & 0.26      & $-$0.10  & 0.29  \\ 
    $x_\mathrm{HI}$      & 0.34    & 0.02     & 0.25     & 0.06      & 0.25     & 0.05  \\
    log $n_\mathrm{H}$   & $-$4.02 & $-$4.40  & $-$3.72  & $-$3.96   & $-$3.62  & $-$3.99 \\
    log $\Delta$         & -0.01   & -0.02    & 0.29     & 0.42      & 0.39     & 0.39   \\
    $\delta_*$           & 0       & 0        & 7.75     & 19.97     & 21.77     & 12.36  \\
    \hline
  \end{tabular}
\end{table}

At $z=7$ (Figure~\ref{fig:ENV_figure_2}), compared to the C II-absorbing gas, C IV-absorbers obviously trace gas that is more abundant in carbon (panel a), has higher $J_\mathrm{912\angstrom}$ (panel c), gas density (panel e), and local average stellar mass (panel f). The H I neutral fraction in panel d for example shows that while this metric is generally quite similar between C II and C IV, C II is arising in a long tail up to 75 per cent, whereas C IV does not for example exceed a neutral fraction of 50 per cent. From panel b of Figure~\ref{fig:ENV_figure_2}, showing the mass fractions of the two transitions, we can see that C II is the preferred ionization state by mass at this redshift of carbon for absorbing cells, with C IV systems arising from cells with only $\sim0<x_\mathrm{CIV}<0.15$.

At $z=5$, the carbon-absorbing gas is still generally more enriched and denser than non-absorbing gas (Table~\ref{tab:property_averages}). However, some global properties have changed. In particular, since reionization has concluded at $z=6$, the UVB amplitude traced by absorbers is more similar to the global value, and the global neutral fraction has dropped to $\sim$0. The ionization fractions of carbon have evolved, with C IV becoming more preferred, although not necessarily more so than C II. Simultaneously, the average conditions tracing C II- and C IV-absorbing gas have converged: they are now arising from more similar gas densities, UVB amplitudes, carbon abundances, and near more similar stellar mass overdensities, although C II is associated with higher $\delta_\mathrm{M_*}$.

It is relevant to note that the cells containing the highest mass fraction of an ion do not necessarily contain the most absorbers in that ion in the simulation. At $z=7$ for example, the global 16th-84th percentile range of $x_\mathrm{CIV}$ extends from $\sim0$ \textendash 10 percent, with a small but extended tail out to 80 percent, but the peak number of absorbers occurs in gas with approximately 4 percent of the mass in C IV. The mean ionization fractions in C II and C IV of the regions containing strong absorbers tend towards the global mean both before and after reionization, rather than preferentially tracing regions with abnormally high fractions of either ion. 

These results help to identify the driver of both the high ionization ions' greater association with galaxies at $z>6$ and the subsequent changes in this relationship. Before the completion of reionization, due to relatively high neutral fractions and consequently higher gas opacities to ionizing photons, metals are preferentially in lower ionization states everywhere except in close proximity to ionized regions around higher stellar mass galaxies (or at least around more galaxies) with higher local UVB amplitudes. These regions also have higher gas densities and, due to the greater overdensity of galaxies and stars, a greater amount of carbon, allowing even gas with a small mass fraction of C IV to create detectable absorbers. Post-reionization, the global decrease in gas opacity and the resulting increase in the mean free path of e.g. C III-ionizing photons allows the higher ionization states to gradually become the preferred ones, especially in more diffuse gas at larger impact parameters. In the case of carbon, this allows gas originally dominated by C II to contain increasingly large amounts of C IV, causing regions that at higher redshifts would be probed by C II to be taken over by C IV, reducing the association between C IV and galaxies.

\begin{figure*}
\includegraphics[width=0.65\textwidth]{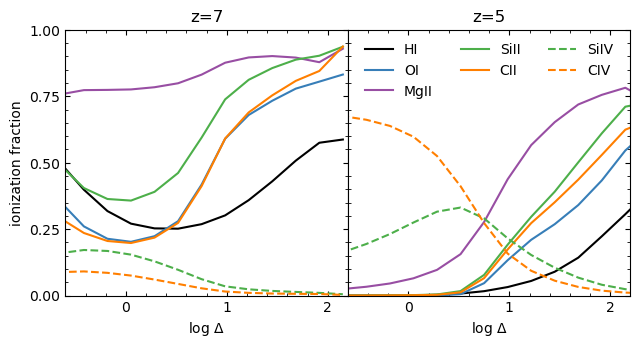}
\caption{Mass-weighted mean ionization fractions of several ions as a function of the local overdensity for $z=7$ and $z=5$. The high ionization states are not preferred for any of these overdensities at $z=7$, but by $z=5$ are preferred for $\log \; \Delta \leq 1$.}\label{fig:ENV_figure_3}
\end{figure*}

To visualize the ionization preferences more explicitly, Figure~\ref{fig:ENV_figure_3} shows the mass-weighted ionization fraction for our ions of interest and H I as a function of the local overdensity in $40$ ckpc voxels, from $\log \Delta = -1$ up to 2.2, i.e. from less than the mean density nearly up to the nominal threshold of the ISM. For $z=7$, the hydrogen neutral fraction decreases from $\sim$55 per cent at $\log \Delta = 2.2$ reaching a nadir of 0.25 at slightly over the mean density before increasing again. All of the low ionization states show similar trends to the hydrogen, with decreases in ionization fraction out to $\sim$ the mean density, below which there is an increase, perhaps due to the filtering of the ionizing background by Lyman limit systems, which results in a spectrally softer background. On the other hand, the high ionization states show gradual increases in ionization fraction as the overdensity decreases. For each low/high ionization element pair, the low-ionization state is universally preferred for all considered overdensities.

By $z=5$, the H I fraction decreases to zero below the CGM threshold at log $\Delta=1$ and the abundance of C IV and Si IV increases. Both C IV and Si IV become preferred over their low ionization counterparts up to the edge of the CGM at log $\Delta\sim1$. Most importantly, we can see that post-reionization, the low ionization ions quickly become disfavored outside of the CGM in terms of mass-weighted percentages.

While high-ionization ions are rare for $z>5.7$ ~\citep{cooper19}, knowledge of these drivers can inform the selection of metal species to use as galaxy tracers at $z\sim5$. We can see that there is a combination of factors driving the galaxy overdensities probed by these ions, in particular, the degree of gas ionization for a given region and the resulting preferred ionization state for an element. However, the overall abundance of an element also plays a role. For example, reionization results in O I being a disfavored ionization state for oxygen in more diffuse gas due to its ionization energy of 1 Ryd. If the neutral fraction and preferred ionization state were the only things driving the nearby galaxy abundance, we would expect O I to be the best galaxy tracer post-reionization; however, Si II is more strongly associated with galaxies. Oxygen is more than an order of magnitude more abundant in the simulation, and so even though O I is rare compared to the other ionization states of oxygen, the overall quantity of O I can still be large enough to produce detectable absorbers in less overdense regions. Generally though, for a given element a lower ionization state will serve as a superior galaxy tracer post-reionization.

\section{Discussion}\label{sec:discussion}
In this work, we have generated mock metal absorption spectra and characterized absorbers' association with galaxies to determine that metal absorption system strengths decrease with increasing impact parameter, and increase with the SFR of the nearest galaxy and the local galaxy overdensity. These trends hold over a range of redshifts $z=5-7$. The local galaxy stellar mass overdensity is also greater around Si IV and C IV at $z=$ 6\textendash7, but decreases post-reionization. In accordance, we find that there are greater numbers of galaxies with $\log L_\mathrm{Ly\alpha}$/(erg s$^{-1}$)$>40.5$ around Si IV and C IV at $z>6$ than at lower redshifts. We generally conclude that (1) metal absorber properties are more dependent on their environment than on the properties of any host galaxy at these redshifts and (2) metal ions probe different physical environments, and further their preferred environment changes with redshift due to the changing physical conditions undergone during hydrogen reionization.

In particular, of interest is that high-ionization absorbers like C IV, itself a common target of high-redshift studies due to its ubiquity and strength, are preferentially associated with lower stellar mass overdensities on 100 pkpc scales post-reionization than several other ions. Conversely at $z>6$, along with Si IV, C IV is associated with higher UVB amplitudes at the Lyman limit, lower neutral fractions, and greater stellar masses. This appears to be due to the completion of reionization, which allows photons capable of ionizing high ionization states to propagate out to larger distances from ionizing sources. Prior to the completion of reionization, C II and other low ionization metal systems can exist at larger distances from galaxies since they do not require a strong source of ionizing radiation, and the simulation has already enriched the regions surrounding galaxies out to large impact parameters. Ultimately, these results show that the ideal targets for galaxy searches post-reionization will be low-ionization metal systems such as Si II rather than high-ionization metals, although there is still a greatly elevated overdensity of galaxies near C IV and Si IV compared to the field. If reionization is not believed to have been completed for a region under survey, high ionization metals may be the preferable tracers for galaxies within a few hundred proper kpc.

\subsection{Comparison to observational studies}
Observational attempts to determine the types of galaxies associated with metal absorbers at $z>5$ have resulted in conflicting conclusions depending on the absorber types. \citet{diaz15} and~\citet{diaz21} searched for LAEs in the environments of known C IV systems from $z=$4.9\textendash5.7 and found an excess of nearby LAE candidates that tended to lie on the fainter end of the LALF, with $L\leq 1.15 L^*_\mathrm{Ly\alpha}$, suggesting a connection due to low mass galaxies' outsized influence on reionization. They concluded that C IV traces the reionized IGM and potentially the sources of ionizing flux density, although it is unclear if any of the detected galaxies could source the metal systems, except in a few cases of systems at very small impact parameters. In contrast, we would conclude that since the conclusion of reionization results in a rapid increase in the mean free path of C III-ionizing photons, corresponding to an increase in the physical locations where C IV might arise, this causes them to no longer preferentially trace the regions with the highest relative ionizing flux density. However, in the context of this simulation, it is difficult to determine the true Ly$\alpha$ luminosities that should be most correlated with high ionization metals given the small box size and the complexity of predicting Ly$\alpha$ emission.

With regards to low ionization metals,~\citet{wu21} recently detected a bright [C II] emitting galaxy near an O I system at $z\sim6$. The inferred halo mass of $4.1\times10^{11} M_\odot$ implies the O I is within the virial radius of the galaxy, at $\sim$ 0.6$R_\mathrm{vir}$. Based on our findings that low ionization metals are at higher overdensities post-reionization, it seems to follow that a low ionization metal like O I would arise quite close to its host galaxy once its surroundings have been reionized. In contrast to our results, however, the star formation rate inferred in~\citet{wu21} is quite high, indeed far higher than for any galaxy in our simulation box.

\begin{figure*}
\includegraphics[width=0.65\textwidth]{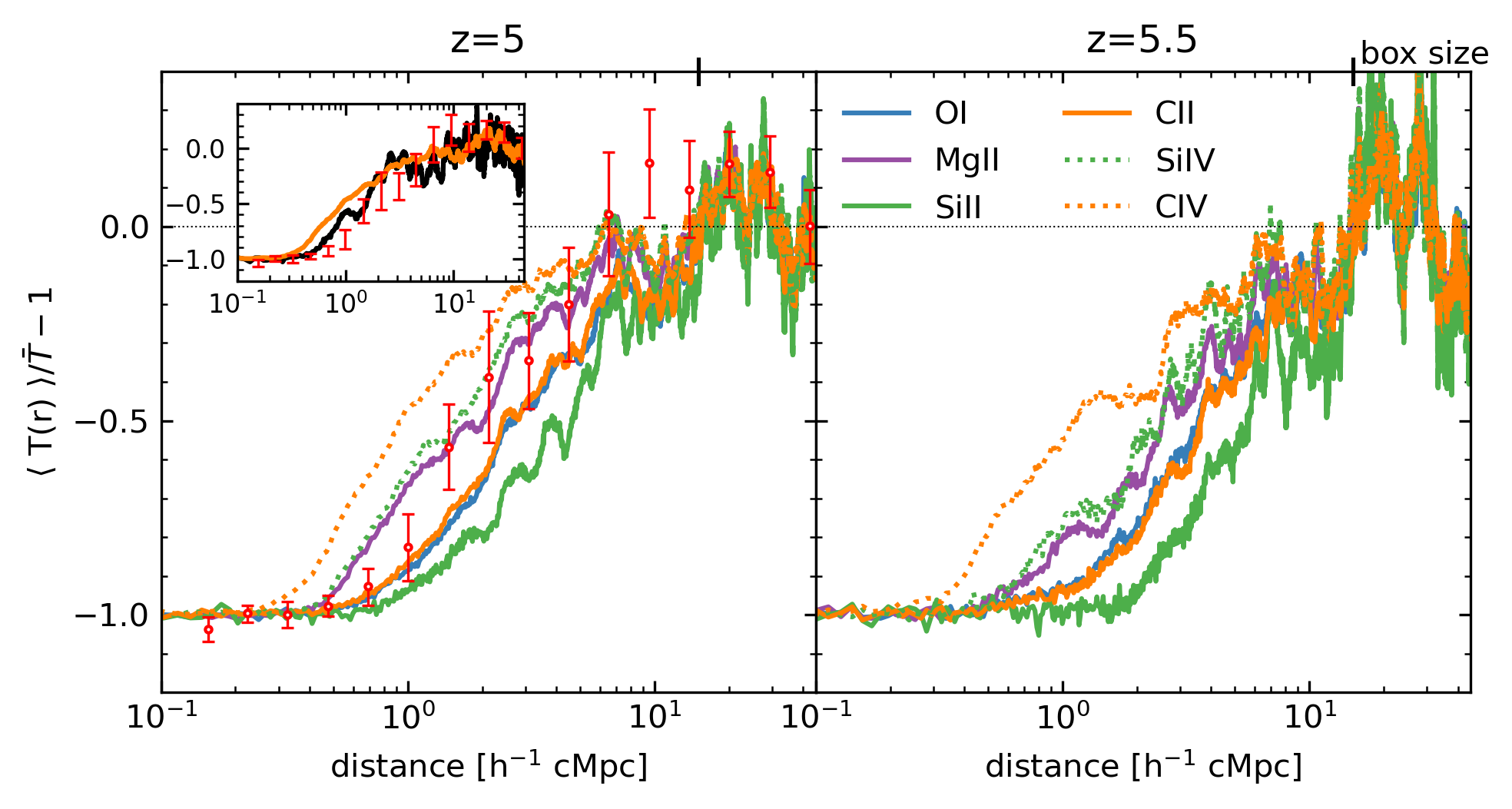}
\caption{The 1D correlation between log N/cm$^{-2}$ $>13$ metal absorption and Ly$\alpha$ transmission at $z=5$. Within 10 cMpc/h, all absorbers generally have less transmission than average, showing their preference to inhabit overdensities. The transmission around high-ionization ions and Mg II are generally higher within this region, likely due to their tendency to arise in less overdense regions in the simulation compared to the low-ionization ions. Results from~\citet{meyer19} at $4.5<z<6.3$ show relatively good agreement for distances beyond 1 cMpc/h, but show greater opacity at smaller distances. The agreement is better for systems with log N/cm$^{-2} >13.7$ (black dashed curve in inset). For a higher redshift of $z=5.5$, the relative positions of the profiles are similar, but the transmission is lower out to larger comoving radii, especially for the lower ionization ions. Note that the distances permitted here exceed the simulation box size, and the correlation values on large scales converge to $\sim$ 0.} \label{fig:DISC_figure_1}
\end{figure*}

\citet{meyer19} examined the 1D correlation between C IV absorption and the Ly$\alpha$ transmission at $4.5 < z < 6.3$, and found greater Ly$\alpha$ opacity within 10 cMpc/h of a C IV system, and enhanced transmission at larger distances. The enhancement in transmission surrounding the C IV systems is interpreted to suggest an excess of ionizing photons provided by clusters of small galaxies. Making an identical comparison with our absorber catalog at $z=5$, we find relatively good agreement with their 1D correlation for $\log N$/cm$^{-2} > 13.5$ (Figure~\ref{fig:DISC_figure_1}), including the greater-than-average transmission at $\sim$ 15 Mpc/h. Reducing the absorber strength cut to 13 to better match their sensitivity limits, we find that the opacity is not as reduced within $r<$ 10 cMpc/h as their measurements, suggesting these absorbers in the simulation are associated with smaller overdensities and thus smaller hosts than in the true $z\sim5$ Universe. This lends further credence to the possibility that simulations tend to expel an excessive amount of metals around small galaxies. Similarly overplotting the Ly$\alpha$ transmissions associated with other ions, we find that for $z=5$ the other ions are associated with regions whose Ly$\alpha$ opacities are greater and either cover a physically larger area or perhaps lie within saturated damped Ly$\alpha$ troughs. Given the correlation between gas overdensity and absorber strengths, we interpret this as further indication that the post-reionization high ion absorbers are located in less overdense regions compared to the other ions, regions which are perhaps occupied by fainter galaxies. These relations at $z=5.5$ show the same general trends remain between the ions, with C IV and Si IV showing higher transmission for a given distance out to $\sim$ 10 Mpc/h. Compared to~\citet{meyer19}, we would similarly conclude that the regions around C IV at $\sim$10 Mpc/h show excess transmission compared to the mean, but note that this is not unique with respect to the regions probed by other metal absorbers.

While it is useful for forecasting purposes to understand the galaxy-absorber association, it is worth considering whether any individual galaxy could be considered the sole contributor of metals to an absorbing system in this epoch, barring cases with very small impact parameters. For very high redshifts, the vast majority of galaxies are quite small and have shallow potential wells with little capability to retain metals~\citep{porciani05}.~\citet{hasan22} related the observed incidence of C IV systems to the abundance of halos hosting galaxies with $L\geq L_*$ and $L\geq 0.1L_*$ out to $z=4.75$, finding that systems with EW$\geq0.04\angstrom$ tend to be too abundant above $z\sim3$ to be hosted by $L_*$ galaxies. However, much fainter galaxies are far too abundant to be the hosts, conceivably suggesting that environments at early times are communally enriched by low-mass galaxies. They determine that a subset of absorbers near massive galaxies, those at $r<0.5 r_\mathrm{vir}$, are strongly physically associated with their host. Given the small size of \emph{Technicolor} galaxies, most of our absorption systems do not fall within the virial radius of the any galaxy, and so would correspond to their sample of weak absorbers tracing a communally enriched medium.

Studies from lower redshifts similarly support a picture in which small satellite galaxies overlie some common metal field, which is further bolstered by our result that the metal absorber strengths are better associated with their local overdensity and environment than with their host traits.~\citet{dutta21} studied low- and high-ionization gas around galaxies at $z<2$ using Mg II and C IV, focusing on correlations between them and the galaxy properties and their environment. They found Mg II strength and incidence to be correlated with galaxy stellar masses and SFRs, with the SFR association being more weak. The covering fraction is found to be higher at a given radius for galaxies in group environments (containing 2 or more galaxies), even when the galaxy is controlled for stellar mass, redshift, and impact parameter. The trends are similar but weaker for C IV.~\citet{anand22} associated Mg II absorbers at $0.4 < z < 1$ with galaxies in cluster environments, looking for correlations between their strength and the traits of the brightest cluster galaxy (BCG) as well as with the nearest satellite galaxies. The Mg II systems are closer on average to satellite galaxies than a random distribution of Mg II, suggesting a physical association. However, the Mg II strengths are not well correlated with the SFR of the nearest galaxy, and their velocity dispersions with respect to the BCG do not match the dark matter kinematics, suggesting they are no longer physically tied to their source galaxies, perhaps due to stripping by the intracluster medium. It seems reasonable to conclude from these that metal absorbers, especially weaker ones at high redshifts, are more likely to be physically unaffiliated with any individual galaxy, and are instead tracing a ``group'' medium.

\subsection{Caveats}\label{ssec:caveats}
Previous analysis of \emph{Technicolor Dawn} has established that while it reproduces the qualitative trends of metal absorption and the approximate order of magnitude of metal line incidence for $z>5$, there are some observational details it cannot reproduce. The column density distributions for all ions are generally too steep, as the simulation generates too many weak metal absorption systems and too few strong ones when compared to observations~\citep{finlator16,finlator18,doughty18,doughty19,hasan20,dodorico22}. The discrepancy is worse for some ions, for example, the underproduction of strong C IV is worse than that of Si IV~\citep[see e.g. Figure 8 of][]{dodorico22}. This tendency is interpreted to mean that the simulation is too efficient at expelling metals from small galaxies, which is further supported by observations finding smaller galaxy-absorber impact parameters than predicted. The detection in~\citet{wu21} of an O I system within 20 pkpc of a relatively bright galaxy is not predicted by the simulation, and the odds of their making a random detection of such a bright galaxy are only $\sim$0.7 per cent. It could also be related to the small box size of the simulation, since only galaxies with SFR $<$ 1 $M_\odot$ yr$^{-1}$ are present at $z\sim6$. The general tendency for stronger absorption systems to be located in more overdense regions in this simulation would suggest that a larger simulation containing larger maximum overdensities should supplement the abundance of strong absorbers to some degree. It is unclear if this could fully alleviate the discrepancy, however, as absorber studies performed with larger simulations still do not agree with observations~\citep[e.g.][]{keating16}.

Another potential issue is our treatment of galaxy Ly$\alpha$ emission, since its transmission approaching the EoR will be drastically reduced. The simulation-based study in~\citet{weinberger19} found that there is attenuation of the redward Ly$\alpha$ peak for intrinsically double-peaked Ly$\alpha$ profiles for a completely reionized Universe, which results in a lower luminosity function even for galaxies brighter than $\log L_\mathrm{Ly\alpha}$/(erg s$^{-1}$)$=43$. If this degree of extinction is suffered by our galaxy sample at $z=6$, then the predicted number of detections for a given luminosity in the fiducial curve in Figure~\ref{fig:OAT_figure_4} would significantly decrease. Further, such a degree of Ly$\alpha$ extinction would indicate that the observed LALFs of~\citet{drake17} and~\citet{delavieuville19} are sampling a galaxy population that is \emph{not} represented in these simulations, and are instead an intrinsically brighter population of galaxies dimmed by external processes. By extension, this would mean that our abundance matching-based L$_\mathrm{Ly\alpha}$ transform does not represent a sensible order-of-magnitude estimate of the detectable galaxy abundance near metals. However, given the generally increasing association of metals with galaxies seen in Figure~\ref{fig:OAT_figure_3}, we expect the general trends and relationships to extend to higher galaxy luminosities.

IGM attenuation is impacted not only by the global gas density and neutral fraction, but by the local degree of ionization surrounding a given galaxy; brighter galaxies residing in larger ionized regions have enhanced transmission fractions when compared with faint galaxies, and thus their observed numbers are less impacted than those of faint galaxies~\citep{weinberger19,park21}. If this is indeed the case, then the main difference in our results is that there would be no evolution with respect to reionization in the types of ions that trace galaxies observable in Ly$\alpha$: low ions would always be the preferential tracer, as long as the H II region surrounding a galaxy was sufficiently large. This would essentially constitute a bias induced by Ly$\alpha$ observability of the galaxy and so would not necessarily hold for other observable galaxy emission, for example [C II] or even the UV magnitude.

Despite this complication, the qualitative result of low ions being the preferred galaxy tracers post-reionization seems robust. There is already observational support for a reduction in the frequency of low ionization absorbers post-reionization, paired with evidence that these systems are preferentially on the low EW and column density side~\citep{becker19,cooper19,doughty19}. These support a model wherein gas at low overdensities transitions from low to high ionization at the conclusion of reionization, increasing the physical association between galaxies and the remaining low ionization gas.

\section{Conclusions}\label{sec:conclusions}
We have used the most recent iteration of the \emph{Technicolor Dawn} simulation to study the association between metal absorbers and galaxies at $z>5$, focusing in particular on the factors that may drive absorber strength, the incidence of galaxies near absorbing systems, and the environmental conditions that drive those associations. At $z=6$ there is a scattered correlation between the SFR of the host galaxy and the column density of the metal system, but there is a stronger and more consistent relationship between the local overdensity of galaxies and the absorber strength. Further, we re-create the observed anti-correlation between metal absorber strength and impact parameter to the host galaxy and the anti-correlation varies in strength between ions, with high-ionization C IV for example showing a profile that is less steep than low-ionization species like O I or Si II. The correlation function of galaxies with high-ionization metal absorbers is elevated compared to other metals for $z=7\rightarrow6$, but the difference diminishes by $z=5$. This translates to high-ionization metals being associated with more galaxies with Ly$\alpha$ luminosities in excess of $10^{40.5}$ erg/s prior to the end of reionization, tuned to occur at $z=6$. This is not restricted to Ly$\alpha$; they are also associated with galaxies with brighter UV magnitudes. Post-reionization this association diminishes and leads to a lower incidence of relatively bright galaxies nearby these specific transitions.

We find that the average environmental conditions tracing metals differs between ionization states, and that these are the conditions that drive the pre-reionization association with galaxies for high-ionization systems. Prior to the end of reionization, high-ionization metal systems arise in regions with higher Lyman limit specific intensities, higher galaxy overdensities, and also higher galaxy stellar mass overdensities. Using C IV as a test case to investigate how this drives their association with galaxies, we find that for $z>6$ C IV absorbers arise in regions with slightly lower minimum H I neutral fractions and higher specific intensities than C II absorbers, but also in regions with a greater carbon abundance and slightly higher gas density. The distribution of neutral fractions and UVB amplitudes for C II extends to extreme tails in high H I neutral fractions and low UVB amplitudes.

These results show that one can use knowledge of the preferred ionization state and overall abundance of an element to select ions which will tend to probe more overdense regions in order to maximize the chance of detecting galaxies in post-reionization high $z$ searches. In particular, absorption systems of low-ionization states of less abundant elements may serve as superior galaxy tracers.

\section*{Acknowledgements}
CD acknowledges helpful conversations with the ENIGMA group at UC Santa Barbara and Leiden University. CD and KF gratefully acknowledge support from STScI Program \#HST-AR-16125.001-A. Support for this program was provided by NASA through a grant from the Space Telescope Science Institute, which is operated by the Associations of Universities for Research in Astronomy, Incorporated, under NASA contract NAS5-26555. Our simulation was run using the Extreme Science and Engineering Discovery Environment (XSEDE), which is supported by National Science Foundation grant number ACI-1548562. The Cosmic Dawn Center is funded by the Danish National Research Foundation under grant No. 140.

\section*{Data Availability}
The data will be shared on reasonable request to the corresponding author.


\bibliographystyle{mnras}
\bibliography{references} 


\bsp	
\label{lastpage}
\end{document}